\documentclass[twocolumn,preprintnumbers, amssymb,amsmath,aps,prd,floatfix,nofootinbib,superscriptaddress,showpacs]{revtex4-2}

\usepackage{epsfig}
\usepackage{amssymb}
\usepackage{amsmath}
\usepackage{lipsum}
\usepackage{graphicx}
\usepackage[dvipsnames]{xcolor}
\usepackage{tikz}
\usepackage{bm}
\usepackage{subfigure}
\usepackage[
    colorlinks,
    linkcolor=blue,
    anchorcolor=black,
    citecolor=blue
]{hyperref}
\usepackage{soul}
\usepackage[version=4]{mhchem}
\allowdisplaybreaks[4]

\begin{document}

\title{Probing Saturation Effect in Heavy Meson Pair Correlations in Forward $pA$ Collisions}

\author{Zhan Gao} \email{gaozhan2@mails.ccnu.edu.cn}
\affiliation{Three Gorges Tourism Polytechnic College, Yichang, Hubei 443111, China}

\author{Cyrille Marquet} 
\email{cyrille.marquet@polytechnique.edu} 
\affiliation{CPHT, CNRS, \'Ecole polytechnique,  Institut Polytechnique de Paris, 91120 Palaiseau, France}

\author{Yu Shi} 
\email{yu.shi@polytechnique.edu}
\affiliation{CPHT, CNRS, \'Ecole polytechnique,  Institut Polytechnique de Paris, 91120 Palaiseau, France}
\affiliation{Key Laboratory of Particle Physics and Particle Irradiation (MOE), Institute of frontier and interdisciplinary science, Shandong University, Qingdao, Shandong 266237, China}

\author{Bo-Wen Xiao}  
\email{xiaobowen@cuhk.edu.cn}
\affiliation{School of Science and Engineering, The Chinese University of Hong Kong (Shenzhen), Longgang, Shenzhen, Guangdong, 518172, P.R. China}
\affiliation{Southern Center for Nuclear-Science Theory (SCNT), Institute of Modern Physics, Chinese Academy of Sciences, Huizhou, Guangdong 516000, China}

\begin{abstract} 
Forward two-particle correlations in $pA$ collisions have long been recognized as a particularly sensitive observable for exploring gluon saturation effects. In the back-to-back regime, two-particle correlations receive substantial contributions from both soft-gluon radiation and saturation effects. In this work, we study heavy meson pair correlations in forward proton–nucleus collisions by incorporating a unified Sudakov resummation for heavy meson pair correlations in the Color Glass Condensate effective theory. Our results are in good agreement with the LHCb $\Delta\phi$ measurements for $D^0\bar D^0$ pairs in forward $pp$ and $pA$ collisions, as well as with the measurements of $J/\psi$ pairs from $b\bar b$ decays in forward $pp$ collisions.  We further present predictions for the $\Delta \phi$ distribution and azimuthal anisotropies of $D\bar D$ and $B\bar B$ correlations in the forward rapidity region at the Large Hadron Collider.  A pronounced mass hierarchy is found in both the nuclear modification factor $R_{pA}$ and the azimuthal anisotropy ratio $\langle \cos 2\phi_{qP} \rangle_{pA}/\langle \cos 2\phi_{qP} \rangle_{pp}$, with stronger suppression for bottom meson pairs than for charm meson pairs. As the rapidity increases, the suppression becomes more pronounced while the mass hierarchy remains robust. This study will help us to search for the saturation signal via heavy-meson pair correlations in forward $pA$ collisions.
\end{abstract}

\maketitle

\section{Introduction}
\label{sec::intro}

In the small-$x$ regime of the high-energy collisions, where $x$ denotes the longitudinal momentum fraction of gluons with respect to their parent nucleon, gluon densities grow rapidly due to Bremsstrahlung radiation, leading to gluon overlap and recombination inside the proton or nucleus. This drives nonlinear evolution and eventual saturation of the gluon fields, forming an ultra-dense gluon matter~\cite{Gribov:1983ivg,Mueller:1985wy,Mueller:1989st,McLerran:1993ni,McLerran:1993ka,McLerran:1994vd}, which is captured by the Color Glass Condensate (CGC) theory~\cite{Gelis:2010nm,Iancu:2003xm}.  This phenomenon is commonly referred to as gluon saturation. The nonlinear evolution behavior of the dense gluon field with decreasing $x$ is governed by the Balitsky-Kovchegov (BK)~\cite{Balitsky:1995ub,Kovchegov:1999yj} and Jalilian-Marian-Iancu-McLerran-Weigert-Leonidov-Kovner (JIMWLK)~\cite{JalilianMarian:1997jx, JalilianMarian:1997gr, Iancu:2000hn, Ferreiro:2001qy} nonlinear evolution equations. A large amount of experimental data from the current high-energy facilities, such as RHIC and the LHC, provides an excellent opportunity to investigate the gluon saturation effect and search for evidence of the ultra-dense gluon matter.  With future high-luminosity colliders, particularly the Electron-Ion Collider~\cite{Boer:2011fh,Accardi:2012qut,AbdulKhalek:2021gbh,AbdulKhalek:2022hcn,Caucal:2026ymj} and the proposed Large Hadron Electron Collider (LHeC), it has become increasingly essential to uncover compelling signatures of the ultra-dense gluon matter. 

The two-particle correlations in the forward region of proton-nucleus ($pA$) collisions, including both light and heavy hadron/jet pairs, have long been viewed as key measurements to probe the ultra-dense gluonic matter at small-$x$. In particular, the azimuthal angular decorrelation of the light hadron pair, firstly calculated in Ref.~\cite{Marquet:2007vb}, has been shown to be highly sensitive to gluon saturation effects, and thus has attracted significant attention from both theoretical and experimental communities.  In particular, in the correlation limit where the pair is nearly back-to-back, a clear hierarchy emerges between the hard scale $\boldsymbol P_h=(\boldsymbol{p}_{1}-\boldsymbol{p}_{2})/2$ and the transverse-momentum imbalance $\boldsymbol q_h=\boldsymbol{p}_{1}+\boldsymbol{p}_{2}$, satisfying $\boldsymbol{P}_h^2 \simeq \boldsymbol{p}_1^2 \simeq \boldsymbol{p}_2^2  \gg \boldsymbol{q}_h^2$. In the Born-level cross section,  the small transverse momentum imbalance is governed by the intrinsic momentum of gluons inside the target, such that the two-particle correlations provide direct sensitivity to the gluon transverse-momentum-dependent (TMD) distribution~\cite{Dominguez:2011wm}. Consequently, studies of forward two-particle correlations not only serve as direct probes of saturation effects, but also provide important insights into the properties of gluon TMDs at small-$x$. Therefore, within the CGC framework, the two-particle production has been extensively studied in $pA$ collisions~\cite{Kharzeev:2004bw,Jalilian-Marian:2005qbq,Marquet:2007vb,Stasto:2011ru,Dominguez:2011wm,Akcakaya:2012si,Jalilian-Marian:2012wwi,Kutak:2012rf,Lappi:2012nh,Rezaeian:2012wa,Stasto:2012ru,Kovner:2014qea,vanHameren:2014ala,Kotko:2015ura,Basso:2015pba,Kovner:2015rna,Basso:2016ulb,Rezaeian:2016szi,vanHameren:2016ftb,Boer:2017xpy,Hagiwara:2017fye,Benic:2017znu,Albacete:2018ruq,Stasto:2018rci,Marquet:2019ltn,vanHameren:2019ysa,Goncalves:2020tvh,Kolbe:2020tlq,vanHameren:2020rqt,Benic:2022ixp,Al-Mashad:2022zbq,Caucal:2025zkl}, and has also broadly investigated at electron-nucleus ($eA$) collisions~\cite{Mueller:2012uf,Mueller:2013wwa,Zheng:2014vka,Altinoluk:2015dpi, Metz:2011wb,Dominguez:2011br,Dumitru:2015gaa,Dumitru:2016jku,Boer:2016fqd,Dumitru:2018kuw,Hatta:2016dxp,Altinoluk:2015dpi,Kotko:2017oxg,Mantysaari:2019csc,Boussarie:2019ero,Salazar:2019ncp,Bergabo:2021woe,Zhao:2021kae,Boussarie:2021ybe,Boer:2021upt,Hagiwara:2021xkf,Iancu:2021rup,Taels:2022tza,Hatta:2022lzj,Caucal:2022ulg,Tong:2022zwp,Caucal:2023fsf,Tong:2023bus,Rodriguez-Aguilar:2023ihz,Caucal:2023nci,Shao:2024nor,Caucal:2024nsb,Shao:2026doo}. 

However, the soft-gluon radiation effect (Sudakov effect) has been shown to induce large logarithmic enhancements in the back-to-back region~\cite{Zheng:2014vka}, which significantly broaden the azimuthal distribution and consequently wash out the saturation signal~\cite{Stasto:2018rci}. Our recent work in Ref.~\cite{Marquet:2025jdr} indicates that the heavy-quark mass effects enhance the sensitivity of heavy-meson pair production to saturation dynamics, resulting in a clear mass-dependent hierarchy in which saturation effects become stronger for $B$-meson pairs than in $D$-meson pairs. In this work, we investigate gluon-initiated processes that are sensitive to three types of gluon TMD distributions.

On the experimental side, the azimuthal angle correlation of $D^0 \bar D ^0$  pair has been measured by the LHCb Collaboration in both forward $pp$~\cite{LHCb:2012aiv} and $p\text{Pb}$~\cite{LHCb:2020jse} collisions. Moreover, the azimuthal correlation of the $J/\psi$ pair originating from the $b \bar b$ decay has been studied in forward $pp$ collisions by the LHCb Collaboration~\cite{LHCb:2017bvf}. Therefore, these measurements of heavy meson pairs in forward $pp$ and $p\text{Pb}$ collisions provide a valuable opportunity to quantitatively explore the saturation effect. 

In the regime $\boldsymbol q^2\ll \boldsymbol P^2$, soft gluon emission effects from massive particles have been extensively studied in both QCD and QED~\cite{Catani:1990eg,Zhu:2013yxa, Klein:2018fmp,Klein:2020jom, delCastillo:2020omr, Ghira:2023bxr, Shi:2024gex, Ghira:2024nkk, Aglietti:2024zhg, Catani:2014qha,Catani:2017tuc,Hatta:2021jcd, Shao:2023zge,Dai:2026hso}.
Notably, in our previous work~\cite{Marquet:2025jdr}, we developed a unified resummation scheme for heavy meson pair correlations in photoproduction that consistently interpolates between the massive and massless limits, thereby ensuring accurate predictions throughout the correlation region. In this work, we extend this framework to $pA$ collisions and apply it to the study of the heavy-meson pair production in forward $pA$ and $pp$ collisions in the back-to-back correlation limit.

Within the CGC framework, for the $\Delta \phi$ distribution, our calculations yield a very good agreement with the LHCb $D^0 \bar D ^0$ data in both $pA$ and $pp$ collisions. They also describe the LHCb data on the $\Delta \phi$, $A_T$, and $\boldsymbol{q}_h$ distributions for $J/\psi$ pairs originating from $b \bar b$ decays in $pp$ collisions. In addition, we further predict $\Delta \phi$ distributions and the asymmetry $\langle \cos 2\phi_{qP} \rangle$ of both $D \bar D$ and  $B \bar B$ correlations in the forward rapidity region.  We find a pronounced mass hierarchy in both the nuclear modification factor $R_{pA}$ and the ratio $\langle \cos 2\phi_{qP} \rangle_{pA}/\langle \cos 2\phi_{qP} \rangle_{pp}$. This indicates that heavy-quark mass effects enhance the sensitivity of heavy-flavor pair correlations to small-$x$  saturation dynamics.

This paper is organized as follows. In Sec.~\ref{sec::res}, we present the theoretical formulas of the resummed improved cross sections within the CGC framework for heavy meson pair production in $pA$ collisions. In Sec.~\ref{sec::set}, we provide the details of the numerical setup used in this work. In Sec.~\ref{sec::num}, we compare our numerical results with LHCb data in the $p\text{Pb}$ and $pp$ collisions, and then present our predictions of both heavy meson pairs at the LHC. Finally, we summarize our findings in Sec.~\ref{sec:con}.

\section{Theoretical Framework}
\label{sec::res}

\begin{figure}[!ht]
\centering
\includegraphics[width=0.4\textwidth]{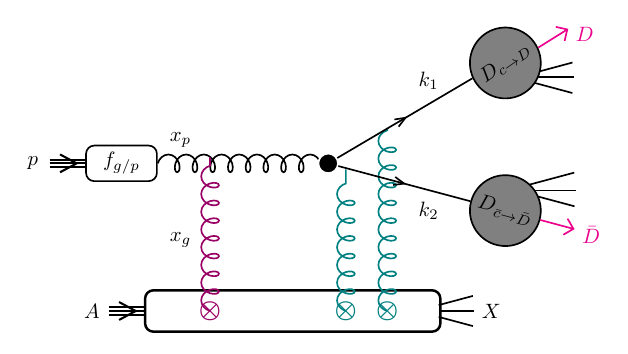}
\caption{Schematic illustration of $D$-meson pair production in $pA$ collisions within the CGC framework, where multiple scattering occurs either before or after the gluon splitting. The purple and teal gluons denote the corresponding initial- and final-state contributions, respectively, arising from distinct amplitude-level diagrams.}
\label{fig:pA}
\end{figure}

In the heavy quark pair production in $pA$ collisions, the dominant contribution arises from the gluon-gluon fusion process, which is given as follows
\begin{equation}
g(x_p)+g(x_g) \rightarrow Q(\boldsymbol{k}_1) + \bar Q (\boldsymbol{k_2}) +X, 
\end{equation}
where the incoming gluon from the proton carries a longitudinal momentum fraction $x_p$, while the gluon from the nucleus is characterized by a longitudinal momentum fraction $x_g$, and the produced heavy quark and antiquark have the transverse momentum $\boldsymbol{k}_1$ and $\boldsymbol{k_2}$, respectively.

In the forward rapidity region $y>0$, where $x_p$ is much larger than $x_g$, the incoming proton can be regarded as a dilute system in which the interaction between the partons can be ignored, and the parton distribution can be described by the collinear parton distribution functions (PDFs). In contrast, the nucleus can be treated as a dense system, where nonlinear gluon dynamics in the dense gluon field become important. In this framework, as shown in Fig.~\ref{fig:pA}, the incoming collinear gluon from the proton undergoes multi-scattering with the dense gluon field inside the target, and subsequently produces the heavy quark pair. The multi-scattering can happen either before or after gluon splitting process.   Within the CGC effective theory, the LO cross-section can be factorized as
\begin{eqnarray}
\frac{d\sigma}{dy_1dy_2 d^2\boldsymbol{k}_1 d^2\boldsymbol{k}_2} \!&=&\!   x_p f(x_p) \mathcal{H}(z,\boldsymbol{P};x_g,\boldsymbol{q}) ,
\end{eqnarray}
where $y_1$ and $y_2$ represent the rapidities of quarks.  $x_p f(x_p)$ denotes the collinear gluon PDF inside the proton. Their total transverse momentum $\boldsymbol{P}$ and transverse momentum imbalance  $\boldsymbol{q}$  are defined as $\boldsymbol{P} = (1-z)\,\boldsymbol{k}_1 - z\,\boldsymbol{k}_2$ and $\boldsymbol{q} = \boldsymbol{k}_1 + \boldsymbol{k}_2$, respectively.  The longitudinal momentum fraction of the heavy quark is given as
\begin{equation}
z=\frac{\sqrt{\boldsymbol{k}_1^{\,2}+m_Q^2}e^{y_1}}{\sqrt{\boldsymbol{k}_1^{\,2}+m_Q^2}e^{y_1}+\sqrt{\boldsymbol{k}_2^{\,2}+m_Q^2}e^{y_2}},
\end{equation}
with heavy-quark mass $m_{Q}$. $\mathcal{H}(z,\boldsymbol{P};x_g,\boldsymbol{q})$ corresponds to the multi-scattering between the collinear gluon splitting process and the dense gluon field~\cite{Blaizot:2004wv,Fujii:2006ab}.
In the back-to-back correlation region ($\boldsymbol{P} \gg \boldsymbol{q}$), the multi-scattering can be factorized as the coupling of the TMD hard factor $\mathcal{H}^{ij}_{\rm TMD}$ and the gluon TMDs $xG_{ij}$~\cite{Dominguez:2011wm,Marquet:2017xwy}. In addition, the soft gluon radiation effect has a significant impact in the back-to-back region, which can be resummed by the Sudakov resummation. 
After including both small-$x$ saturation effect and the soft gluon radiation effect, the cross-section of the heavy meson pair correlation can be factorized as
\begin{eqnarray}
\frac{d\sigma}{d \Omega} \!&=&\!\int  x_p f(x_p) \otimes \mathbf{} D_{h_1/Q} (z_{1h}) \otimes D_{h_2/\bar Q}(z_{2h})e^{i\boldsymbol{q}\cdot\boldsymbol{b}}   \nonumber \\
&& \!
\otimes\mathcal{H}^{ij}_{\rm TMD}(z,\boldsymbol{P})\otimes xG_{ij}(x_g, \boldsymbol{b}) \otimes \mathcal{S}(m_Q,Q,\boldsymbol{b}),
\end{eqnarray}
with $d\Omega=dy_{1}\,dy_{2}\,d^{2}\boldsymbol{p}_{1}\,d^{2}\boldsymbol{p}_2$ and fragmentation function $D_{h_i/Q} (z_{ih})$. $\boldsymbol{p}_{1}=z_{1h}\boldsymbol{k}_{1}$ and $\boldsymbol{p}_{2}=z_{2h}\boldsymbol{k}_{2}$ denote the transverse momenta of heavy mesons with rapidities $y_1$ and $y_2$. The invariant mass of the final quark pair is defined as $Q^2=(k_1+k_2)^2\simeq\left[2+2\cosh(y_1-y_2)\right]\boldsymbol{P}^2$. $x_{p}=(\sqrt{\boldsymbol{k}_1^{\,2}+m_Q^2}e^{ y_{1}}+\sqrt{\boldsymbol{k}_2^{\,2}+m_Q^2}e^{ y_{2}})/\sqrt{s}$ is the longitudinal momentum fraction of dilute gluon in the proton, while $x_{g}=(\sqrt{\boldsymbol{k}_1^{\,2}+m_Q^2}e^{ -y_{1}}+\sqrt{\boldsymbol{k}_2^{\,2}+m_Q^2}e^{ -y_{2}})/\sqrt{s}$ represents the longitudinal momentum fraction of dense gluon in the nucleus with the collisional energy $\sqrt{s}$.
These gluon TMDs are defined as~\cite{Mulders:2000sh,Boer:2010zf,Dominguez:2010xd, Metz:2011wb,Dominguez:2011wm,Marquet:2016cgx,Marquet:2017xwy}
\begin{eqnarray}
&\!\! xG_{(1)}^{ij}(x_{g},\boldsymbol{q}) =\int {\mathcal{P.S.}}  \langle{\rm Tr}\left[(\partial_{i}U_{y}) ( \partial_j U_{x}^{\dagger})\right] {\rm Tr}\left[U_{x}U_{y}^{\dagger}\right]\rangle ,\\
&\!\! xG_{(2)}^{ij}(x_{g},\boldsymbol{q}) =- \int {\mathcal{P.S.}} \langle{\rm Tr}\left[(\partial_{i}U_{x})U_{y}^{\dagger} \right]
{\rm Tr}\left[ (\partial_{i}U_{y})U_{x}^{\dagger}\right]\rangle,\\
&\!\! xG_{(3)}^{ij}(x_{g},\boldsymbol{q}) =-N_c \int {\mathcal{P.S.}} \langle{\rm Tr}\left[(\partial_{i}U_{x})U_{y}^{\dagger}(\partial_{i}U_{y})U_{x}^{\dagger}\right]\rangle, \\
\!\!&\!\! xG_{(ADP)}^{ij}(x_{g},\boldsymbol{q}) =xG_{(1)}^{ij}(x_{g},\boldsymbol{q}) - xG_{(2)}^{ij}(x_{g},\boldsymbol{q}),
\end{eqnarray}
with $\int {\mathcal{P.S.}}= 4/(g_s^2 N_c)/ (2\pi)^{3} \int d^{2}\boldsymbol{x}d^{2}\boldsymbol{y}\exp\left[-i\boldsymbol{k}\cdot(\boldsymbol{x}-\boldsymbol{y})\right]$. $ U_x$ denotes the Wilson line in the fundamental representation. $ xG_{(ADP)}^{ij}$ denotes the dipole-type gluon TMD in the adjoint representation~\cite{Marquet:2017xwy}.
In the coordinate space, these gluon TMDs are defined as
\begin{equation}
xG^{ij}(x_{g},\boldsymbol{b}) =\int d^2\boldsymbol{q} e^{i\boldsymbol{q}\cdot \boldsymbol{b}} xG^{ij}(x_{g},\boldsymbol{q}).
\end{equation}
These three gluon TMDs can be decomposed into two terms, the unpolarized component and the linearly polarized component, which are defined as~\cite{Metz:2011wb, Dominguez:2011br, Marquet:2016cgx,Marquet:2017xwy}
\begin{equation}
xG_{(m)}^{ij}(x_{g},\boldsymbol{q})
 = \frac{\delta^{ij}}{2} {xf}_{gg}^{(m)}(x_{g},\boldsymbol{q}) + \Pi^{ij}(\boldsymbol{q}){xh}_{gg}^{(m)}(x_{g},\boldsymbol{q}).
\end{equation}
where the traceless rank-two projector is defined as $\Pi^{ij}(\boldsymbol{q})= (\boldsymbol{q}^i \boldsymbol{q}^j)/\boldsymbol{q}^2- \delta^{ij}/2$, whose contraction with $2  \boldsymbol{P}^i \boldsymbol{P}^j \Pi^{ij}(\boldsymbol{q})/\boldsymbol{P}^2=\cos2\phi_{qP}$.

The soft function $\mathcal{S}(m_Q,Q,\boldsymbol{b})$ encodes the multi-pole soft gluon radiation effects from the initial dilute and dense gluons, and the final heavy quarks. The standard approach is to perform analytic resummation of these soft gluon radiation contributions. In the resummation framework, the soft function $\mathcal{S}(m_Q,Q,\boldsymbol{b})$ can be expressed as
\begin{eqnarray}
\mathcal{S}(m_Q,Q,\boldsymbol{b}) = \exp\left[-{\rm Sud}(\boldsymbol{b}, m_Q,Q) \right] .
\end{eqnarray}
 For heavy meson pair production in forward $pA$ and $pp$ collisions, the resummation-improved cross section is given by
\begin{widetext}
\begin{align}
\frac{d\sigma}{d\Omega} 
&= \int \frac{dz_{1h}}{z_{1h}^2} \int \frac{dz_{2h}}{z_{2h}^2} 
\int\frac{d^{2}\boldsymbol{b} }{(2\pi)^{2}}e^{i\boldsymbol{q}\cdot\boldsymbol{b}}e^{-{\rm Sud}(\boldsymbol{b}, m_Q,Q)}   D_{h_1/Q}(z_{1h})D_{ h_2/\bar Q}(z_{2h})  \left[ 1 -C_{2}(\mu_b) \frac{2C_F \alpha_s(\mu_b)}{\pi} \cos(2\phi_{bP}) \right]
\nonumber \\
 & \times \frac{\alpha_{s}^2z(1-z)}{2C_F} x_p g(x_p, \mu_b)
 \int d^2 \boldsymbol{q}^\prime e^{-i\boldsymbol{q}^\prime \cdot \boldsymbol{b}} \Bigg\{
\mathcal H^{f}_{\rm TMD}(z,\boldsymbol{P}) \left[xf_{gg}^{(1)}(x_{g},\boldsymbol{q}^\prime) -  2z(1-z) xf_{ADP}(x_{g},\boldsymbol{q}^\prime) 
-\frac{1}{N_c^2} xf_{gg}^{(3)}(x_{g},\boldsymbol{q}^\prime)  \right] \nonumber \\
 &\quad \quad+\cos(2\phi_{Pq^\prime})\mathcal H^{h}_{\rm TMD}(z,\boldsymbol{P})
\left[xh_{gg}^{(1)}(x_{g},\boldsymbol{q}^\prime) -  2z(1-z) xh_{ADP}(x_{g},\boldsymbol{q}^\prime) 
-\frac{1}{N_c^2} xh_{gg}^{(3)}(x_{g},\boldsymbol{q}^\prime)  \right]
\Bigg\}
 .
\label{eq::DDbar}
\end{align}
The Sudakov effects contain the perturbative and NP components, which are given by
\begin{eqnarray}
{\rm Sud}(\boldsymbol{b}, m_Q,Q) ={\rm Sud}_{\rm per}(\boldsymbol{b},m_Q,Q)  + {\rm Sud}_{\rm NP}(\boldsymbol{b},Q).
\label{eq::Sud}
\end{eqnarray}
The perturbative Sudakov factor includes contributions from both incoming dilute and dense gluons, as well as the outgoing heavy quarks. The heavy-quark Sudakov factor in the gluon–gluon fusion channel is the same as that in the photon–gluon fusion process. For the latter, the corresponding double-logarithmic contribution was computed in our previous work~\cite{Marquet:2025jdr}, while the single-logarithmic contribution was derived in Ref.~\cite{Shao:2026doo}. For the symmetric rapidity configuration $y_1=y_2$, the perturbative Sudakov factor is given by
\begin{equation}
{\rm Sud}_{\rm per} = C_A\,{\rm Sud}^{i} + 2\,C_F \,  {\rm Sud}^Q_{f},
\end{equation}
with
 \begin{align}
 \!{\rm Sud}^{i}=&\int ^Q _{\mu_b} \frac{d\mu }{\mu } \frac{\alpha_s (\mu)}{\pi}\left( \ln \frac{\boldsymbol{P}^2}{\mu^2}-2 \beta_0 
 \right) 
  +
 \int ^Q _{\mu_b} \frac{d\mu }{\mu } \frac{\alpha_s (\mu)}{\pi} \left(
  \ln \frac{Q^2}{\mu^2}-2 \beta_0 
 \right) ,
 \label{eq::iniSu}
 \\
 \!\!{\rm Sud}^Q_{f}\!=&\!\int ^Q _{\mu_b} \frac{d\mu }{\mu } \frac{\alpha_s(\mu) }{\pi} \left[ \ln \frac{Q^2}{\mu^2}\! - \! \ln \frac{m_Q^2}{\mu^2}  \theta\left( m_Q-\mu \right)\theta\left( m_Q-\mu_b \right) \right] + \frac{N_c}{C_F} \frac{N_c-2C_F}{4 C_F- N_c} \int ^Q _{\mu_b} \frac{d\mu }{\mu } \frac{\alpha_s(\mu) }{\pi}  \ln \frac{Q^2}{\boldsymbol{P}^2}\nonumber\\
 &
 -\int ^Q _{\mu_b} \frac{d\mu }{\mu } \frac{\alpha_s(\mu) }{\pi} \left[  \frac{3}{2}- \frac{1}{2}\theta\left( m_Q-\mu \right)\theta\left( m_Q-\mu_b \right) \right]
 ,
\label{eq::QSu}
 \end{align}
 with $\mu_b^2=4e^{-2 \gamma_E}/\boldsymbol{b}^2$. In Eq.~(\ref{eq::QSu}), the first term represents the heavy-quark double-logarithmic contribution, while the last term corresponds to the heavy-quark single-logarithmic contribution. The second term was first derived for $gg\rightarrow q\bar q$ jet production in Ref.~\cite{Hatta:2021jcd}. In the regime $\mu_{b}^{2}<m_Q^2$, at double-logarithmic accuracy, the initial and final contributions agree with the results obtained for the jet case~\cite {Hatta:2021jcd} after making the replacement $m_Q^2\rightarrow \boldsymbol{P}^2 R^2$. In addition, we include the Sudakov single-logarithmic contributions from the initial-state gluons on both the dense and dilute sides, as well as from the final-state heavy-quark pair. For heavy-quark pair photoproduction, the heavy-quark Sudakov single-logarithmic contribution has been studied in diffractive processes in Ref.~\cite{Shao:2026doo}.

Here, we include the single logarithmic contribution associated with the
initial-state gluons, while the single logarithmic contribution from the
final-state heavy-quark pair is not included in the present treatment. 
Accordingly, the factorization scale entering the PDF is set to $\mu_b$.

For the symmetric rapidity configuration $y_1=y_2$, the harmonic coefficient $C_{2}(\mu_{b})$ is given by
\begin{align}
C_{2}(\mu_{b})=&\ln\frac{M_{{\rm in}}^{2}}{\mu_{b,2}^{2}}-\frac{m_{Q}^{2}\boldsymbol{b}^{2}}{8}\,_{2}F_{3}\left(1,1;2,2,3;-\frac{m_{Q}^{2}\boldsymbol{b}^{2}}{4}\right)+\ln\frac{1}{4}-1+\frac{1}{2C_F N_{c}} \frac{2+N_c^2}{4N_c C_F-N_c^2} (1-\ln4) ,
\label{eq::c2}
\end{align}
where $_{2}F_{3}$ denotes the generalized hypergeometric function and $\ln\mu_{b,2}^{2}=\ln\mu_{b}^{2}+2\psi^{(0)}(1)+2\gamma_E+1$. Apart from the last term,  the harmonic coefficient $C_{2}(\mu_{b})$   coincides with the corresponding coefficient in photon-induced process~\cite{Marquet:2025jdr}. In the region $\mu_{b,2}^{2}<m_Q^2$, Eq.~(\ref{eq::c2}) matches the jet result for $gg\rightarrow q\bar q$ in Eq.~(74) of Ref.~\cite {Hatta:2021jcd}, upon the replacement $m_Q^2\rightarrow \boldsymbol{P}^2 R^2$. The study of the corresponding higher asymmetries $\langle \cos2n \phi_{qP} \rangle$ is left for future work.  

The TMD hard factors $\mathcal H^{(i)}_{\rm TMD} $ are given as~\cite{Marquet:2017xwy}
\begin{eqnarray}
 \!\mathcal H^{f}_{\rm TMD} (z,\boldsymbol{P}) \! &=&  \! \frac{\left(\boldsymbol{P}^{2}+m_Q^{2}\right)^{2} \mathcal{P}_{qg}(z) +2z(1-z) m_Q^{2}\boldsymbol{P}^{2}}{\left(\boldsymbol{P}^{2}+m_Q^{2}\right)^{4}},\quad \text{and}
 \qquad
 \!\mathcal H^{h}_{\rm TMD} (z,\boldsymbol{P}) =   z(1-z)\frac{ 2m_Q^{2}\boldsymbol{P}^{2}}{\left(\boldsymbol{P}^{2}+m_Q^{2}\right)^{4}}.
\end{eqnarray}
with $\mathcal{P}_{qg}(z)= [z^2+(1-z)^2]/2$.
\end{widetext}

\section{Numerical setup}
\label{sec::set}
In this section, we discuss the numerical setup used in this work. In the Gaussian approximation and large-$N_c$ limit, the relevant gluon TMDs can be expressed in terms of the quark dipole amplitude $S(x_g,\boldsymbol{r})$ and its Fourier transform $F(x_g,\boldsymbol{q})$. In this work, we obtain the dipole amplitude by solving the rcBK equation and reconstruct the gluon TMDs within this approximation, rather than performing full JIMWLK evolution. The corrections beyond the large-$N_c$ limit and Gaussian approximation can be quantified using full JIMWLK evolution. For studies of the Gaussian approximation to JIMWLK evolution and its limitations, as well as finite-$N_c$ corrections to gluon TMDs, see Refs.~\cite{Kovchegov:2008mk,Marquet:2010cf,Dumitru:2011vk,Iancu:2011nj,Cougoulic:2026drr,Lappi:2026kxi}. In the Gaussian approximation and large-$N_c$ limit, these unpolarized gluon TMDs shown in Eq.~(\ref{eq::DDbar}) are given by~\cite{Dominguez:2011wm,vanHameren:2016ftb,Marquet:2017xwy}
\begin{eqnarray}
\!\!xf_{gg}^{(1)}(x_{g},\boldsymbol{q}) &=&\mathcal{K}_1 \int d^2 \boldsymbol{l} \boldsymbol{l}^2 F(x_g,\boldsymbol{l}) F(x_g, \boldsymbol{q}-\boldsymbol{l}), \\
\!\!xf_{gg}^{(2)}(x_{g},\boldsymbol{q})\!\!& =&\!\! -\mathcal{K}_1 \! \int \!\! d^2 \boldsymbol{l} (\boldsymbol{q}-\boldsymbol{l})\cdot \boldsymbol{l}  F(x_g,\boldsymbol{l}) F(x_g, \boldsymbol{q}-\boldsymbol{l}), \nonumber\\ \\
\!\!xf_{gg}^{(3)}(x_{g},\boldsymbol{q}) \!\! &=& \!\! \mathcal{K}_2 \int \!\!\frac{d^2\boldsymbol{r}}{\boldsymbol{r}^2}e^{-i\boldsymbol{q} \cdot \boldsymbol{r}} \!\!\left[1\!\!-\!\left( S(x_{g},\boldsymbol{r}) \right)^2\!\right], \label{eq:f3} \\
xf_{\tiny{ADP}}(x_{g},\boldsymbol{q})  &=&xf_{gg}^{(1)}(x_{g},\boldsymbol{q})-xf_{gg}^{(2)}(x_{g},\boldsymbol{q}) \nonumber \\
& =&\mathcal{K}_1 \frac{\boldsymbol{q}^2 }{2} \int d^2 \boldsymbol{l}  F(x_g,\boldsymbol{l}) F(x_g, \boldsymbol{q}-\boldsymbol{l}), \label{eq::fadf}
\end{eqnarray}
and these linearly polarized gluon TMDs are listed below~\cite{Taels:2017shj,Marquet:2017xwy}
\begin{align}
xh_{gg}^{(1)}(x_g, \bm{q}) =& \mathcal{K}_1\int d^2\bm{l}  \frac{2(\bm{q} \cdot \bm{l})^2 -\bm{q}^2 \bm{l}^2}{\bm{q}^2} F(x_g, \bm{l}) F(x_g, \bm{q} - \bm{l}),\label{eq::h1} \\
x h_{ADP}(x_g,\bm{q})=& xf_{ADP}(x_g,\bm{q}),\label{eq::hadp} \\
x h_{gg}^{(3)}(x_g, \bm{q}) =& \mathcal{K}_2 \int \frac{d^2\boldsymbol{r}}{\boldsymbol{r}^2}\frac{J_2(|\boldsymbol{q} || \boldsymbol{r}|)}{2\ln \left(\frac{1}{\boldsymbol{r}\Lambda} + e\right)} \left[ 1 - (S(x_g, \boldsymbol{r}))^2 \right].\label{eq::h3}
\end{align}
with $\mathcal{K}_1=\frac{N_c S_\perp}{2\pi^2 \alpha_s}$ and $\mathcal{K}_2= \frac{C_F}{\pi^2 N_c}\mathcal{K}_1$. 
$S_\perp$ is the transverse area of the target. $xh_{gg}^{(1)}$ and $xh_{ADP}$ are given in Eq.~(F.41) and (F.42) of Ref.~\cite{Taels:2017shj}. For the Weizs\"{a}cker-Williams (WW) distributions $xf_{gg}^{(3)}$ and $xh_{gg}^{(3)}$, in our implementation, we took an additional approximation inherited from the MV-model expression: a more accurate calculation within the Gaussian approximation would evaluate the derivative structure in Eq.~(50) of Ref.~\cite{Marquet:2017xwy} using the evolved dipole amplitude. Instead, we have replaced those by their MV expressions: $1/\boldsymbol{r}^2$ in $xf_{gg}^{(3)}$ and $1/[\boldsymbol{r}^2 \ln \left(\frac{1}{\boldsymbol{r}\Lambda} + e\right) ]$ in $xh_{gg}^{(3)}$. We have compared this simplification with a fuller calculation using derivatives of the evolved dipole amplitude and found negligible impact to the cross-section. This is due in part to the $1/N_c^2$ suppression of those WW distributions as shown in Eq.~\ref{eq::DDbar}; in addition the $xh_{gg}^{(3)}$ contribution is further suppressed by the Sudakov effect.
 
 The dipole $S$-matrix and $F(x_g,\boldsymbol{q})$ encode the non-linear small-$x$ evolution and saturation dynamics, which can be described by the BK/JIMWLK evolution equations.
$F(x_g,\boldsymbol{q})$ is typically defined as the Fourier transform of the quark dipole amplitude $S(x_{g},\boldsymbol{r})$, and can be written as
\begin{align}
  F(x_{g},\boldsymbol{q})  =  \int \frac{d^2\boldsymbol{r}}{(2\pi)^2}e^{-i\boldsymbol{q}\cdot\boldsymbol{r}} S(x_{g},\boldsymbol{r}) , \label{eq:f} 
\end{align}
with
\begin{eqnarray}
S(x_{g},\boldsymbol{r}=\boldsymbol{x}-\boldsymbol{y}) =\frac{1}{N_c} \langle {\rm Tr} U_{x} U^\dagger _{y}  \rangle.
\end{eqnarray}
In this work, we obtain the dipole amplitude $S(x_g,\boldsymbol{r})$ by solving the running-coupling BK (rcBK)  evolution equation~\cite{Golec-Biernat:2001dqn,Balitsky:2006wa, Kovchegov:2006vj, Gardi:2006rp, Balitsky:2007feb,Albacete:2007yr, Albacete:2010sy,Berger:2010sh}, which is given by 
\begin{eqnarray}
\frac{d N(Y, \boldsymbol{r})}{d Y} =& \int d^2 \boldsymbol{r}_1  \mathcal{K}_{\text{BK}}(\boldsymbol{r}, \boldsymbol{r}_1, \boldsymbol{r}_2) 
 [ N(Y, \boldsymbol{r}_1)+ N(Y, \boldsymbol{r}_2) \nonumber
\\
&-N(Y, \boldsymbol{r}) - N(Y, \boldsymbol{r}_1)N(Y, \boldsymbol{r}_2) ],
\label{eq:bk}    
\end{eqnarray}
with $N(Y,\boldsymbol{r}) = 1 - S(Y,\boldsymbol{r})$, $\boldsymbol{r}_2= \boldsymbol{r}-\boldsymbol{r}_1$ and  $Y = \ln(x_0/x_g)$. To incorporate running-coupling effects, we adopt the Balitsky prescription~\cite{Balitsky:2006wa}, under which the evolution kernel takes the form
\begin{align}
     \mathcal{K}_{\mathrm{BK}}(\boldsymbol{r}, \boldsymbol{r}_1, \boldsymbol{r}_2)=&\frac{\alpha_s(\boldsymbol{r})N_c}{2\pi^2}\left[\frac{\boldsymbol{r}^2}{\boldsymbol{r}_1^2\boldsymbol{r}_2^2}+\frac{\alpha_s(\boldsymbol{r}_1)-\alpha_s(\boldsymbol{r}_2)}{\alpha_s(\boldsymbol{r}_2)}\frac{1}{\boldsymbol{r}_1^2}\right. \nonumber \\&\left.+\frac{\alpha_s(\boldsymbol{r}_2)-\alpha_s(\boldsymbol{r}_1)}{\alpha_s(\boldsymbol{r}_1)}\frac{1}{\boldsymbol{r}_2^2}\right].   
\end{align}

In this study, we use the modified McLerran-Venugopalan (MV) model as the initial input for rcBK evolution, which is given by~\cite{Albacete:2010sy}
\begin{equation}
S(x_{0}=0.01,\boldsymbol{r})=\exp\left[-\frac{\left(Q_{s0}^2 \boldsymbol{r}^2 \right)^{\gamma} }{4}\ln\left(\frac{1}{|\boldsymbol{r}|\Lambda}+e\right)\right].
\label{eq::S0}
\end{equation}
with $\Lambda = 0.24\, \rm{GeV}$. We use the parameter set with anomalous dimension $\gamma = 1.074$ and the saturation scale for the proton $Q_{s0}^2 = 0.15\,  \mathrm{ GeV^2}$, obtained from DIS fits including heavy-quark contributions~\cite{Albacete:2010sy}.  For the nucleus, we adopt $Q_{s0,A}^2 = 5Q_{s0,p}^2$ in  Eq.~(\ref{eq::S0}). To estimate the uncertainty associated with the initial condition, we also present results obtained with $Q_{s0,A}^2 = 3Q_{s0,p}^2$ and compare them with the default choice, thereby illustrating the impact of varying the nuclear saturation scale. 

In the large $\boldsymbol{q}$ regime, performing the numerical Fourier transform of the Bessel function appearing in Eqs.~(\ref{eq:f3})  and (\ref{eq:f}) to obtain these gluon TMDs is challenging, and leads to numerical instabilities. On the analytical side,  $xf_{gg}^{(3)}(x_g, \boldsymbol{q})$ and $F(x_g, \boldsymbol{q})$ in Eqs.~(\ref{eq:f3})  and (\ref{eq:f}) exhibit a power-law analytical behavior at large $\boldsymbol{q}$ regime with $\boldsymbol{q} ^2 \gg Q_s^2$. Therefore, we adopt the power-law fit method developed in Ref.~\cite{Shi:2021hwx} to capture the asymptotic behavior of $xf_{gg}^{(3)}(x_g, \boldsymbol{q})$ and $F(x_g, \boldsymbol{q})$ at large $\boldsymbol{q}$ regime.

\begin{figure}[!h]
\includegraphics[width=0.7\linewidth]{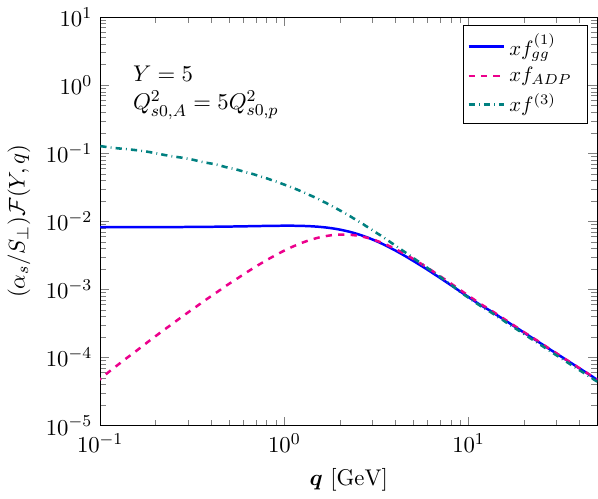}
\caption{
The unpolarized gluon TMDs ($xf_{gg}^{(1)}$, $xf_{gg}^{(3)}$, and $xf_{ADP}$) obtained from the rcBK evolution equation as a function of $\boldsymbol q$ for $Q_{s0, A}^2 = 5Q_{s0,p}^2$ at $Y=5$.
}
\label{fig:Fgg}
\end{figure} 

Fig.~\ref{fig:Fgg} shows the unpolarized gluon TMDs ($xf_{gg}^{(1)}$, $xf_{gg}^{(3)}$, and $xf_{ADP}$) obtained from the rcBK evolution as functions of $q$ for $Q_{s0, A}^2 = 5Q_{s0,p}^2$ at $Y=5$.  In the low $\boldsymbol{q}$ region, where $\boldsymbol{q}^2\ll Q_s^2$, these unpolarized gluon TMDs exhibit distinct behaviors: as $\boldsymbol{q}$ decreases, the $xf_{gg}^{(1)}$ remains nearly flat, the $xf_{gg}^{(3)}$ increases slowly, while $xf_{ADP}$ decreases due to the prefactor $\boldsymbol{q}^2$ in Eq.~(\ref{eq::fadf}). This indicates that saturation effects, non-linear evolution dynamics, lead to different behaviors among these unpolarized gluon TMDs. In contrast, in the high $\boldsymbol{q}$ region, where $\boldsymbol{q}^2 \gg Q_s^2$, all three unpolarized  TMDs tend to coincide and exhibit a universal power-law behavior $\sim 1/\boldsymbol{q}^2$.

We adopt the CTEQ (CT18) PDFs parametrizations~\cite{Hou:2019efy} for the collinear gluon distribution $g(x_p,\mu^2)$.  The $c$ and $b$ quarks hadronisation into $D$ and $B$ mesons is described employing the Peterson fragmentation functions (FFs)~\cite{Peterson:1982ak}. The Peterson parametrizations do not include the collinear single logarithms and are independent of $\mu_F$. Therefore, the final-state Sudakov single logarithms can be included directly, following a treatment analogous to that used for diffractive heavy-quark pair production at the parton level in Ref.~\cite{Shao:2026doo}.

To estimate the uncertainty associated with the choice of FFs, we also employ the KKKS FFs for  $c\rightarrow D$ fragmentation~\cite{Kniehl:2006mw,Kneesch:2007ey}  and KKSS FFs for $b\rightarrow  B$ fragmentation~\cite{Kniehl:2007erq}. For using the KKKS or KKSS FFs, these FFs resum part of the collinear logarithms $\alpha_s\ln(\mu_F^2/m_Q^2)$, while the corresponding logarithmic contributions from the region $\mu_F^2<m_Q^2$ are not included. Thus, the collinear logarithmic resummation is incomplete over the full $\mu_F$ range. This has an important consequence for the choice of the FF factorization scale. In particular, when using the KKKS or KKSS FFs, it is not consistent to set $\mu_F=\mu_b$ throughout the full $b$-space integration. For final-state heavy-quark fragmentation, part of the Sudakov single-logarithmic contribution is intertwined with the collinear logarithms resummed through the fragmentation functions, and these effects should ideally be treated consistently. Since such a complete treatment is not available in these two FF parameterizations, we refrain from including the heavy-quark final-state Sudakov single logarithms. Instead, we vary the fragmentation scale within $\mu_F\in[m_Q,Q]$ for the KKKS and KKSS FFs, which probes part of the residual uncertainty associated with this incomplete treatment. The size of the full neglected final-state single-logarithmic contribution, however, cannot be reliably quantified. For the initial-state gluon, in contrast, it is well known that the Sudakov single logarithms can be consistently combined with the resummation of the collinear logarithms through PDF evolution, leading to the natural choice of the gluon-PDF factorization scale $\mu_F=\mu_b$.

For the hadronization of a $b$ quark into a $J/\psi$ meson, following Ref.~\cite{Kniehl:1999vf, Bolzoni:2013tca}, we construct an effective FF describing the transition of $b \rightarrow J/\psi+X$ via an intermediate $B$ meson, expressed as a convolution  given by
\begin{eqnarray}
\!D_{b \rightarrow J/\psi}(z, \mu)=\!\int ^1_z \!\frac{dx}{x} D_{B/b}(z/x, \mu) \frac{1}{\Gamma_B} \frac{d\Gamma}{dx}(x, P_B),\label{eq::b_jpsi_FFs}
\end{eqnarray}
where $\Gamma_B$ denotes the  total decay width of the $B$ meson, and ${d\Gamma}(x, P_B)/{dx}$ is the differential decay distribution for $B \rightarrow J/\psi+X$, with the momentum fraction $x$. The three-momentum  of $B$ meson is given by $\boldsymbol{P}_B^2=[p^2_{J/\psi,T}+(p^2_{J/\psi,T}+m_b^2)\sinh^2y]/x^2$ with the $J/\psi$ transverse momentum $p^2_{J/\psi,T}$. The expressions for $\Gamma_B$ and ${d\Gamma}(x, P_B)/{dx}$ are provided in Ref.~\cite{Kniehl:1999vf}.

In our numerical calculations, we set the heavy-quark masses to $m_b = 4.5$ GeV for the $b$-quark and $m_c = 1.2$ GeV for the $c$-quark. Following Ref.~\cite{Shi:2021hwx}, we take $S^{pA}_{\perp}=1830$ mb for $pPb$ collisions, and $S^{pp}_{\perp}=51$ mb for $pp$ collisions.
The NP Sudakov factor in Eq.~(\ref{eq::Sud}) is parametrized as
\begin{eqnarray}
{\rm Sud}_{\rm NP}(\boldsymbol{b},Q) =
\mathcal{C}_{\rm NP} \left(
0.016\,\boldsymbol{b}^2+0.21\ln \frac{Q}{Q_0}\ln \frac{\boldsymbol{b}^2}{b_*^2} \right), 
\end{eqnarray}
with $Q^2_0=2.4$ GeV$^2$~\cite{Sun:2014dqm}. Here, we follow our previous work in the $\gamma A$ collisions~\cite{Marquet:2025jdr} and use the NP free parameter $\mathcal{C}_{\rm NP}$ to account for the NP from the initial gluon and the final-state heavy quark pair. In the photon-induced process, $\mathcal{C}_{\rm NP}=1$ was found to provide a good description of the H1 
$\Delta\phi$ data for $D$-jet correlations~\cite{Marquet:2025jdr}. Compared with $\gamma+g\to Q\bar Q$, the present $g+g\to Q\bar Q$ channel contains one additional initial-state gluon. Since the expression in the parentheses is adopted from a fit to Drell--Yan data~\cite{Sun:2014dqm}, where the NP Sudakov effect is mainly associated with quark-dominated channels, we rescale the corresponding gluon contribution by the color-factor ratio $C_A/C_F$. Accordingly, we take
\begin{equation}
\mathcal{C}_{\rm NP}=1 + C_A/C_F,
\end{equation}
where the first term represents NP contribution present in the photon-induced baseline~\cite{Marquet:2025jdr}, while the factor $C_A/C_F$ accounts for the NP effect from the additional initial collinear gluon.This is a phenomenological ansatz motivated by the color factors, whose value has not been determined through a dedicated fit to data. This ansatz yields good agreement with the data presented below.

The $b_{*}$ prescription is adopted in the work, defined as $b_{\star} =b_{\perp}/\sqrt{1+b_{\perp}^2/b_{\max}^2}$  with $b_{\max}=1.5~\mathrm{GeV^{-1}}$~ \cite{Davies:1984sp,Ladinsky:1993zn,Landry:2002ix,Prokudin:2015ysa,Sun:2014dqm}. In the perturbative Sudakov factor,  $b$ is consistently replaced by $b_*$. The one-loop running coupling effect is also considered in this work.

\section{Numerical results}
\label{sec::num}

In this section, we compare our numerical results with LHCb data in the $p\text{Pb}$ and $pp$ collisions, and then present our predictions of both heavy meson pairs at the LHC.  For the calculations in this section, the bands indicate the scale uncertainties from varying the fragmentation scale $\mu_F$ of the KKKS/KKSS FFs within $[m_Q, Q]$. Note that the Peterson FFs are independent of $\mu_F$, and therefore its results are represented by single lines. For the comparison between the KKKS/KKSS and Peterson FFs from Subsection~\ref{subsec::phi} to subsection~\ref{subsec::cos}, we do not include the final-state heavy-quark Sudakov single-logarithmic contributions, in order to ensure a consistent comparison. Then, in Subsection~\ref{subsec::sl}, we study the impact of the final-state heavy-quark single-logarithmic contribution using the Peterson FFs. Finally, in subsection~\ref{subsec::Qs}, we investigate the dependence on the saturation scale $Q_{sA}^2$ including the heavy-quark single-logarithmic contributions.

\subsection{The $\Delta \phi$ distribution}
\label{subsec::phi}

Firstly, we will discuss the calculations of the $\Delta \phi$ distribution for the heavy meson pair production.  $\Delta \phi$ denotes the angle difference of the two mesons, defined by $\Delta \phi = |\phi_{\mathbf{p_1}}-\phi_{\mathbf{p_2}}|$. The self-normalized $\Delta \phi$ distribution is given by
\begin{equation}
\frac{1 }{\sigma}\frac{d\sigma}{d \Delta \phi} = \frac{\pi/2}{\sigma} \int dy_{1}dy_{2}d \boldsymbol{p}_{1}^2 d \boldsymbol{p}_2 ^2  \frac{d\sigma}{dy_{1}dy_{2}d^{2}\boldsymbol{p}_{1}\,d^{2}\boldsymbol{p}_2},
\end{equation}
where $\sigma$ denotes the cross-section integrated over the region $\phi \in [0.8\pi,\pi]$. We emphasize that our formalism is valid only in the back-to-back correlation limit, where the transverse momentum imbalance is small and $\Delta\phi$ is close to $\pi$, and therefore our analysis focuses on the away-side region. 
To better illustrate the gluon saturation effect, we also present the nuclear modification factor  differential in $\Delta\phi$, defined by 
\begin{equation}
    R_{pA}(\Delta\phi) = \frac{1}{A} \frac{d\sigma_{pA}/d\Delta\phi}{d\sigma_{pp}/d\Delta\phi},
\end{equation} 
with the nuclear number $A$. We present the numerical predictions of the $\Delta \phi$ distribution for heavy meson pairs, based on the resummation-improved cross-section as shown in Eq.~(\ref{eq::DDbar}), without the final-state heavy-quark single logarithmic contribution.

\begin{figure}[!ht]
\includegraphics[width=0.75\linewidth]{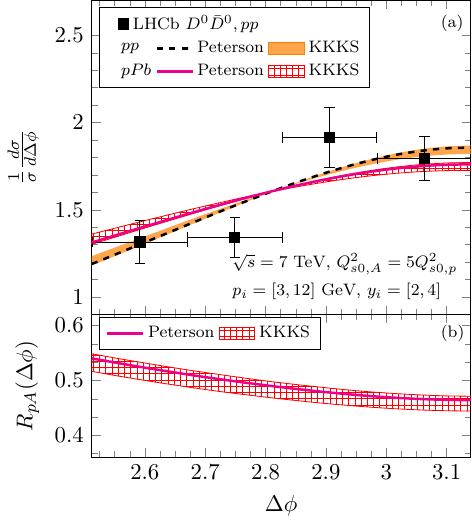}
\includegraphics[width=0.75\linewidth]{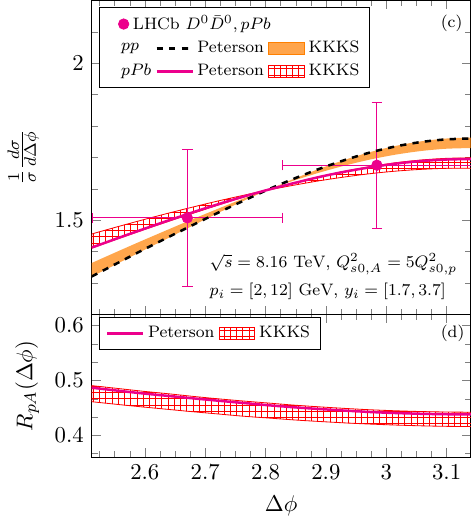}
\caption{
Comparison of theoretical predictions with LHCb data for the $D^0\bar{D}^0$ $\Delta\phi$ distribution in $pp$~\cite{LHCb:2012aiv} (upper panel) and $p\text{Pb}$~\cite{LHCb:2020jse} (lower panel) collisions, together with the corresponding nuclear modification factor $R_{pA}$.
}
\label{fig:dphi_FFs}
\end{figure} 

Panels $(a)$ and $(c)$ of Fig.~\ref{fig:dphi_FFs} show the distributions of the azimuthal angle difference $\Delta \phi = |\phi_{\mathbf{p_1}}-\phi_{\mathbf{p_2}}|$ of final-state $D^0\bar{D}^0$ pair, compared with the LHCb data in $pp$ collisions at $\sqrt s =7$ TeV~\cite{LHCb:2012aiv} and in $pPb$ collisions at $8.16$ TeV~\cite{LHCb:2020jse}. We focus on the away-side distributions, $[0.8\pi,\pi]$, and present self-normalized distributions for both theoretical results and the experimental data within this interval. For comparison, we present corresponding theoretical calculations for $pPb$  and $pp$ collisions at the same center-of-mass energy and with the same kinematic cuts. The magenta solid lines represent the Peterson results for $pPb$ collisions, while the black dashed lines correspond to the Peterson results for $pp$ collisions. The corresponding KKKS results are denoted by the red and orange shaded bands, respectively. As shown in the panels $(a)$ and $(c)$ of Fig.~\ref{fig:dphi_FFs},  our results can provide a very good description of the LHCb experimental data in both $pp$~\cite{LHCb:2012aiv} and $pPb$~\cite{LHCb:2020jse} collisions. The results obtained with the KKKS FF, including its small uncertainty band, are very close to those obtained with the Peterson FF. Moreover, at both collision energies, the $pPb$ results exhibit a stronger $p_\perp$ broadening effect compared to $pp$, which originates from stronger saturation momentum $Q_s^2$ in the nucleus. 

To further illustrate saturation effects, panels $(b)$ and $ (d)$ of Fig.~\ref{fig:dphi_FFs} show $R_{pA}$ as a function of $\Delta\phi$. The $R_{pA}$ shows a strong suppression near the back-to-back region ($\Delta \phi \sim \pi$) and increases gradually as $\Delta \phi$ moves away from this region.

\begin{figure}[!ht]
\includegraphics[width=0.75\linewidth]{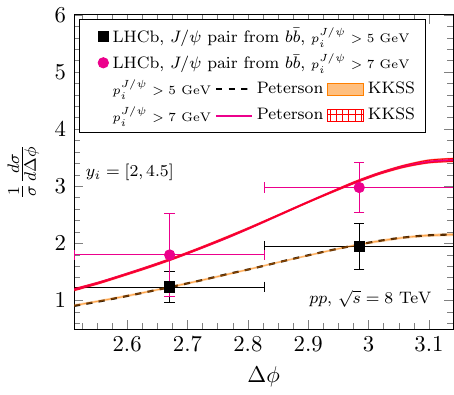}
\caption{
Comparison between theoretical predictions and LHCb data for the azimuthal angle $\Delta\phi$ distribution of the $J/\psi$ meson pair decaying from $b\bar{b}$ pair~\cite{LHCb:2017bvf} in forward $pp$ collisions. The results of $p^{J/\psi}_i>7$ GeV are multiplied by 1.5 for better visibility.
}
\label{fig:bbar_to_Jpsi_pair}
\end{figure}

Fig.~\ref{fig:bbar_to_Jpsi_pair} shows comparisons between theoretical predictions and LHCb data for the self-normalized $\Delta\phi$ distribution of the $J/\psi$ meson pair decaying from a $b\bar{b}$ pair~\cite{LHCb:2017bvf} in forward $pp$ collisions on the away side $\Delta\phi\in [0.8\pi,\pi]$. The LHCb data measure the $b$-hadron pair correlations reconstructed via their inclusive decays into a $J/\psi$ pair in the forward rapidity region $2<y<4.5$.
In our theoretical calculations, we evaluate the non-prompt $J/\psi$ pair production from the $b\bar b$ pair. For the hadronization of a $b$ quark into a $J/\psi$ meson,  we employ an effective FF describing the transition of $b \rightarrow J/\psi+X$, as defined in Eq.~(\ref{eq::b_jpsi_FFs}). 
The magenta solid (black dashed) lines represent the results obtained with the Peterson FFs for $p^{J/\psi}_i >$ 5(7) GeV, while the corresponding KKSS results are shown by the red (orange) shaded bands.
Our numerical results can provide a good description of the LHCb experimental data in both cases. As shown in Fig.~\ref{fig:bbar_to_Jpsi_pair}, the azimuthal angular correlation 
in the back-to-back region becomes steeper for higher transverse momentum 
particles because, although the Sudakov effect grows with $p_T$ and contributes 
to angular decorrelation, its magnitude increases more slowly than the hard 
scale $p_T$ itself. As a result, higher-$p_T$ particle pairs ($p^{J/\psi}_i>7$ GeV) remain more 
collimated near $\Delta\phi \sim \pi$, producing a narrower and sharper 
correlation peak compared to their lower-$p_T$ counterparts ($p^{J/\psi}_i>5$ GeV). The KKSS and Peterson FFs yield very similar results, with only a very small uncertainty band from the fragmentation-scale variation.

\begin{figure}[!ht]
\includegraphics[width=0.75\linewidth]{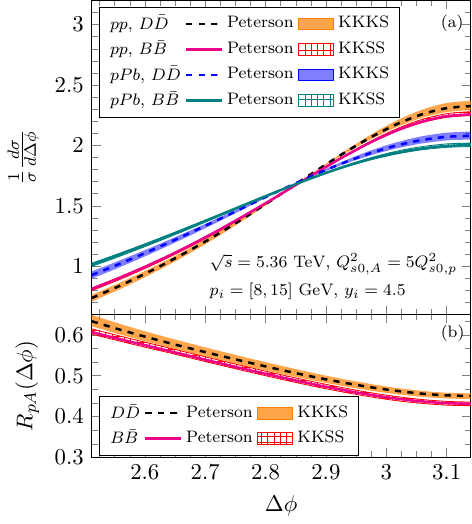}
\includegraphics[width=0.75\linewidth]{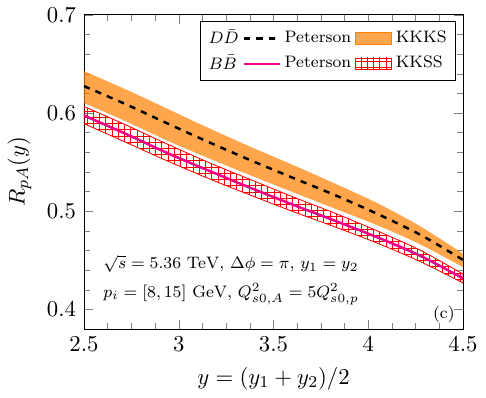}
\caption{
Upper panel: Predictions of the normalized $\Delta\phi$ distributions of $D\bar{D}$ and $B\bar{B}$ pairs in very forward $pp$ and $pPb$ collisions at $y_i = 4.5$ and $p_i \in [8, 15]$ GeV, together with the corresponding nuclear modification factor $R_{pA}$. Lower panel: $R_{pA}$ as a function of rapidity $y$ for $D\bar{D}$ and $B\bar{B}$ pairs at $\Delta\phi=\pi$.
}
\label{fig:DB5360}
\end{figure}

In panel $(a)$ of Fig.~\ref{fig:DB5360}, we present predictions of the $\Delta\phi$ distribution of $D\bar{D}$ and $B\bar{B}$ pairs in very forward $pp$ and $pPb$ collisions at $\sqrt{s} = 5.36$ TeV, with $p_i \in [8, 15]$ GeV and rapidity $y_i = 4.5$. The blue (black) dashed lines represent the $D\bar{D}$ results using the Peterson FFs in $pPb$ ($pp$) collisions, while the corresponding KKKS results are shown by the blue (orange) bands. Similarly, the teal (magenta) solid lines correspond to the $B\bar{B}$ results using the Peterson FFs in $pPb$ ($pp$) collisions, with the KKSS results given by the teal (red) bands. For both $D\bar{D}$ and $B\bar{B}$ pairs in the very forward region, the transverse-momentum ($p_\perp$) broadening effect is particularly strong and clearly visible, where angular correlations in $pPb$ are significantly broader than in $pp$ collisions.

Comparing  $D\bar{D}$ results and $B\bar{B}$ results, we find that the $B\bar{B}$ distribution is broader than the $D\bar{D}$ one in contrast to the behavior observed in $\gamma A$ collisions~\cite{Marquet:2025jdr}. This difference originates from the contribution of the initial-state Sudakov factor in $pA$ collisions, which is twice as large as in $\gamma A$ collisions~\cite {Marquet:2025jdr}, thereby reducing the relative importance of final-state mass effects. As a result, the Sudakov effect in $B\bar B$ production is comparable to that in $c\bar c$ production. 

However, the heavy-quark mass suppresses the fragmentation process, leading to a broadening of the distribution. The larger average fragmentation fraction for $B$ mesons ($\langle z_b \rangle \sim 0.8$) compared to $D$ mesons ($\langle z_c \rangle \sim 0.6$) implies smaller effective parent-quark momenta for $b$ quarks than for $c$ quarks in the same meson kinematics $p_i \in [8,15]$ GeV, corresponding to approximately $p_i \in [10,19]$ GeV and $p_i \in [13,25]$ GeV, respectively. Since distributions associated with smaller parton momenta are typically smoother, this results in a broader $\Delta \phi$ distribution for $B \bar B$ pairs compared to $D \bar D$ pairs. 

Furthermore,  a smaller momentum of the heavier $b$ quarks corresponds to a smaller gluonic momentum fraction $x_g$, which leads to a larger saturation scale $Q_s$.  Consequently, the $R_{pA}$ of $B\bar B$ is more suppressed than that of $D\bar D$, as shown in the panel $(b)$ of Fig.~\ref{fig:DB5360}. Overall, the nuclear modification factor of heavy meson pairs shows a clear mass hierarchy, $R_{pA}\big|_{m_b}<R_{pA}\big|_{m_c}$, indicating enhanced sensitivity to saturation effects at small $x$.

In addition, as the rapidity $y$ increases, the gluon momentum fraction $x_g$ decreases, leading to stronger suppression in $R_{pA}$ at more forward rapidities. This behavior is clearly observed in our predictions shown in panel $(c)$ of Fig.~\ref{fig:DB5360} for both $D \bar D$ and $B \bar B$ pairs, where $R_{pA}$ decreases monotonically with rapidity $y$.
Meanwhile, the mass hierarchy remains robust across the entire forward rapidity range.

\subsection{The $q_h$ and $A_T$ distribution}

\begin{figure}[!ht]
\includegraphics[width=0.78\linewidth]{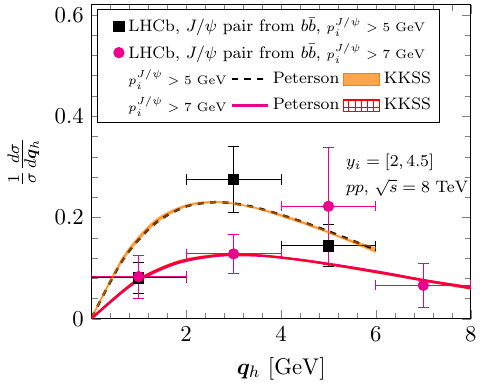}
\caption{
Comparison of our theoretical predictions with LHCb data for the $\boldsymbol{q}_h$ distribution of the $J/\psi$ meson pair decaying from a $b\bar{b}$ pair~\cite{LHCb:2017bvf} in forward $pp$ collisions.
}
\label{fig:bbar_to_Jpsi_pair_qt}
\end{figure}

Next, we present the numerical results of the $\boldsymbol{q}_h$ distribution, which is given by 
\begin{equation}
\frac{d\sigma}{d \boldsymbol{q}_h} = 2\pi \int dy_{1}dy_{2}d\boldsymbol{P}_h d\phi_{qP} |\boldsymbol{P}_h| |\boldsymbol{q}_h|  \frac{d\sigma}{dy_{1}dy_{2} d^2 \boldsymbol{q}_h d^2\boldsymbol{P}_h},
\end{equation}
where $\boldsymbol{q}_h$ characterizes the transverse momentum imbalance between two mesons. The normalization factor for the normalized $\boldsymbol{q}_h$ distribution is given by the cross section integrated over the region $\boldsymbol{q}_h=[0,4]$ GeV, where $\boldsymbol{q}_h< \boldsymbol p_i ^{J/\psi}$. Fig.~\ref{fig:bbar_to_Jpsi_pair_qt} shows comparisons between theoretical predictions and LHCb data for the self-normalized $\boldsymbol{q}_h$ distribution of the $J/\psi$ meson pair decaying from $b\bar{b}$ pair~\cite{LHCb:2017bvf} in forward $pp$ collisions.  The result for $p^{J/\psi}_i>7$ GeV shows stronger broadening than the $p^{J/\psi}_i>5$ GeV result, and our calculation describes the data well for both cuts.

\begin{figure}[!ht]
\includegraphics[width=0.78\linewidth]{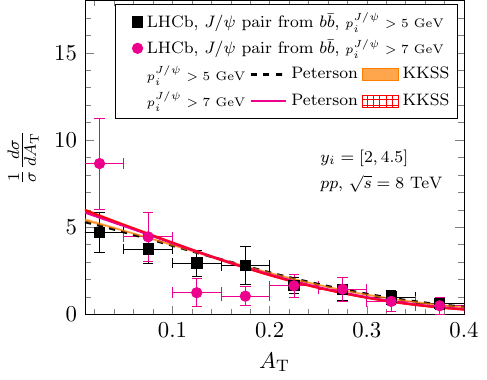}
\caption{
Comparison of our theoretical predictions with LHCb data for the $A_T$ distribution of the $J/\psi$ meson pair decaying from a $b\bar{b}$ pair~\cite{LHCb:2017bvf} in forward $pp$ collisions.
}
\label{fig:bbar_to_Jpsi_pair_AT}
\end{figure} 

We also present the numerical results of the $A_T$ distribution, which is given by
\begin{equation}
\frac{d\sigma}{d A_T} = \int dy_{1}dy_{2}d^2 \boldsymbol{p}_{1} d ^2\boldsymbol{p}_2  \delta \left(A_T-\Bigg|\frac{|\boldsymbol{p}_{1}|-|\boldsymbol{p}_{2}|}{|\boldsymbol{p}_{1}|+|\boldsymbol{p}_{2}|}\Bigg| \right) \frac{d\sigma}{d\Omega},
\end{equation}
where $A_T$ characterizes the transverse momentum between two mesons and is given as $A_T=\Big|\frac{|\boldsymbol{p}_{1}|-|\boldsymbol{p}_{2}|}{|\boldsymbol{p}_{1}|+|\boldsymbol{p}_{2}|}\Big|$.  Fig.~\ref{fig:bbar_to_Jpsi_pair_AT} shows a comparison of our theoretical predictions and LHCb data for the self-normalized $A_T$ distribution of the $J/\psi$ meson pair decaying from $b\bar{b}$ pair~\cite{LHCb:2017bvf} in forward $pp$ collisions.  The result for $p^{J/\psi}_i>7$ GeV is very similar to that for $p^{J/\psi}_i>5$ GeV. Our numerical calculations provide a good description of the LHCb experimental data in both cases.

For both $A_T$ and $\boldsymbol{q}_h$ distributions, the results obtained with the KKSS FF are close to those based on the Peterson FF, and the uncertainty band from varying the fragmentation scale is very small.

\subsection{The asymmetry $\langle \cos 2\phi_{qP} \rangle$}
\label{subsec::cos}

\begin{figure}[!ht]
\includegraphics[width=0.75\linewidth]{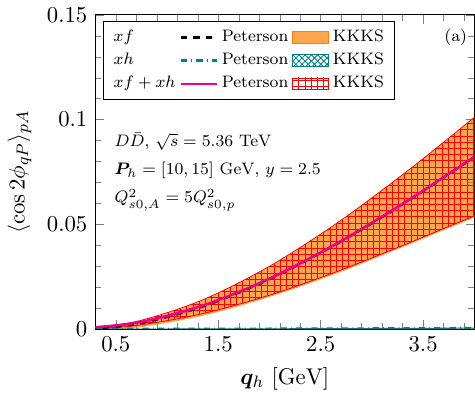}
\includegraphics[width=0.75\linewidth]{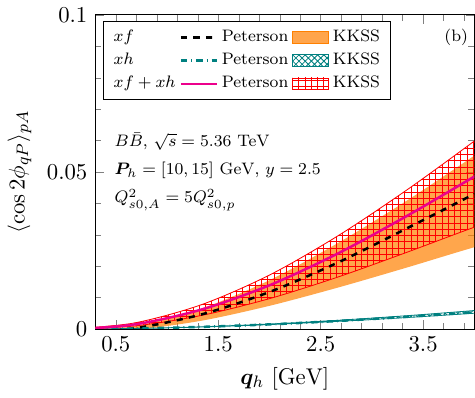}
\caption{Predictions for the $\langle\cos 2\phi_{qP}\rangle_{pA}$ asymmetry as a function of $q_h$ for the $D\bar{D}$ pair (a) and $B\bar{B}$ pair (b) in $pPb$ collisions at $\sqrt{s} = 5.36~\text{TeV}$. The calculations are evaluated at $y = 2.5$ with $P_h \in [10, 15]~\text{GeV}$. The individual contributions from the unpolarized gluon TMD $xf$ and the linearly polarized gluon TMD $xh$, as well as their sum $xf+xh$, are presented.
}
\label{fig:DB5360_cos2phi_xfxh}
\end{figure}

\begin{figure}[!ht]
\includegraphics[width=0.75\linewidth]{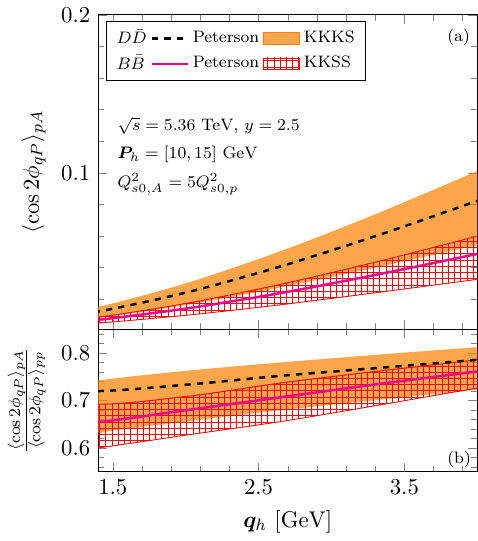}
\\
\includegraphics[width=0.75\linewidth]{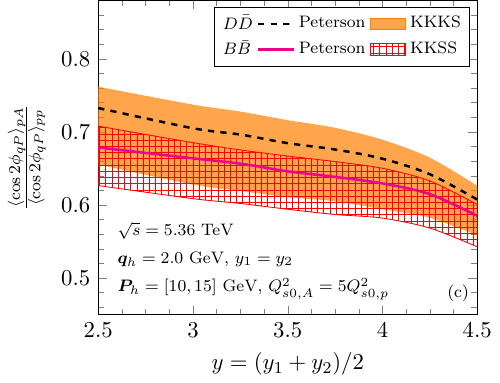}
\caption{
Upper panel: Theoretical results for the $\langle\cos 2\phi\rangle$ asymmetries of $D\bar{D}$ and $B\bar{B}$ pairs in very forward $pPb$ collisions at $y_i = 2.5$ and $\bm{P}_h \in [10, 15]$ GeV, along with the ratio of these asymmetries between $pPb$ and $pp$ collisions. Lower panel: The ratio $\frac{\langle \cos 2\phi_{qP} \rangle_{pA}}{\langle \cos 2\phi_{qP} \rangle_{pp}}$ as a function of $y$ for $D\bar{D}$ and $B\bar{B}$ pairs calculated at $\bm{q}_h=2$ GeV.
}
\label{fig:DB5360_cos2phi}
\end{figure}

Finally, we present the numerical results of the asymmetry $\langle \cos 2\phi_{qP} \rangle$, which is given by 
\begin{equation}
\langle \cos 2\phi_{qP} \rangle =  \frac{\int dy_{1}dy_{2}d\boldsymbol{P}_h d\phi_{qP} |\boldsymbol{P}_h| |\boldsymbol{q}_h| \cos (2\phi_{qP})  \frac{d\sigma}{d\Omega}}{\int dy_{1}dy_{2}d\boldsymbol{P}_h d\phi_{qP} |\boldsymbol{P}_h| |\boldsymbol{q}_h|  \frac{d\sigma}{d\Omega}},
\end{equation}
where $\phi_{qP}$ is defined as azimuthal angle between $\boldsymbol P_h$ and $\boldsymbol q_h$. In the following, we present the numerical predictions for $\langle \cos 2\phi_{qP} \rangle$ using the
resummation-improved cross section given in Eq.~(\ref{eq::DDbar}).

As shown in Fig.~\ref{fig:DB5360_cos2phi_xfxh}, we present the $\langle \cos 2\phi_{qP} \rangle$ predictions from different gluon TMDs contributions for $D\bar{D}$ and $B\bar{B}$ pairs in very forward $pPb$ collisions at $\sqrt{s} = 5.36$ TeV, with $\boldsymbol{P}_h \in [10, 15]$ GeV and rapidity $y_i = 2.5$. Since the contribution from the linearly polarized gluon TMD $x h$ is
proportional to $m_Q^2$, it is much smaller for $D\bar D$ pairs than for
$B\bar B$ pairs. In addition, the $x h$ contribution is further suppressed by
the Sudakov effect. As a result, the dominant contribution arises from the
coupling between the unpolarized gluon TMD $x f$ and soft-gluon radiation.
A similar behavior is also observed in other processes~\cite{Boer:2017xpy,Zhao:2021kae,Marquet:2025jdr}. Finally, we find sizable asymmetries for both $D$ and $B$ meson pairs.

In panel $(a)$ of Fig.~\ref{fig:DB5360_cos2phi}, we present predictions of $\langle \cos 2\phi_{qP} \rangle$ of $D\bar{D}$ and $B\bar{B}$ pairs. The $D\bar{D}$ distribution is stronger than the $B\bar{B}$ one, similar to the behavior observed in $\gamma A$ collisions~\cite{Marquet:2025jdr}. This mass-ordering effect originates from the coefficient $C_2 \propto \ln (Q^2/m_Q^2)$.  The results obtained with the KKKS and KKSS FFs show a clear scale dependence when the fragmentation scale is varied from $m_Q$ to $Q$. The results from Peterson FF lie within the uncertainty band generated by this scale variation.

To better illustrate the gluon saturation effect, we also present the $\langle \cos 2\phi_{qP} \rangle$ ratios between $pA$ and $pp$ collisions, defined by 
\begin{equation}
 \frac{\langle \cos 2\phi_{qP} \rangle_{pA}}{\langle \cos 2\phi_{qP} \rangle_{pp}}.
\end{equation} 
As shown in panel $(b)$, the ratio $\frac{\langle \cos 2\phi_{qP} \rangle_{pA}}{\langle \cos 2\phi_{qP} \rangle_{pp}}$ is more strongly suppressed for $B\bar B$ pairs than that of $D\bar D$ pairs, indicating that heavy quark mass effects enhanced sensitivity to saturation dynamics. Moreover, as the rapidity $y$ increases, the suppression becomes stronger for both $D \bar D$ and $B \bar B$ pairs, while the clean mass-ordering pattern remains robust. This trend is clearly observed in panel $(c)$ of Fig.~\ref{fig:DB5360_cos2phi}.

\subsection{Impact of Final-State Heavy-Quark Single-Logarithmic Contributions}
\label{subsec::sl}

\begin{figure}[!ht]
\includegraphics[width=0.7\linewidth]{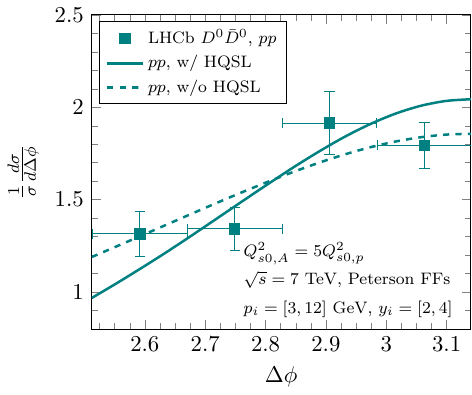}
\includegraphics[width=0.7\linewidth]{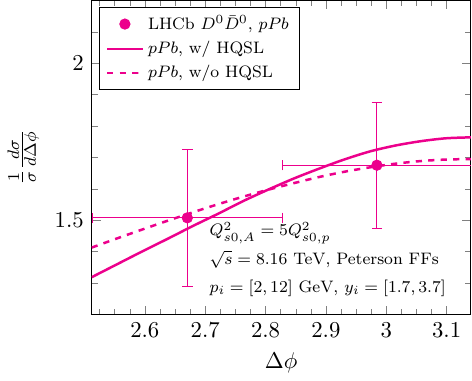}
\caption{
Comparison of theoretical predictions with LHCb data for the $D^0\bar{D}^0$ azimuthal  $\Delta\phi$ distributions in $pp$~\cite{LHCb:2012aiv} (upper panel) and $p\text{Pb}$~\cite{LHCb:2020jse} (lower panel) collisions, illustrating the impact of the single-logarithmic Sudakov contribution from heavy quarks.
}
\label{fig:dphi_DD_HQSL}
\end{figure} 

\begin{figure}[!ht]
\includegraphics[width=0.7\linewidth]{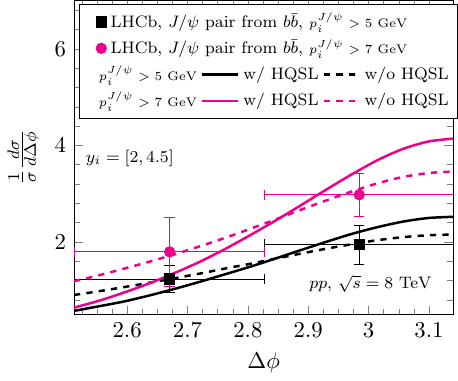}
\includegraphics[width=0.7\linewidth]{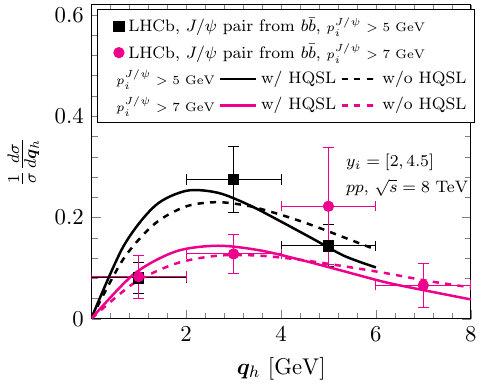}
\includegraphics[width=0.7\linewidth]{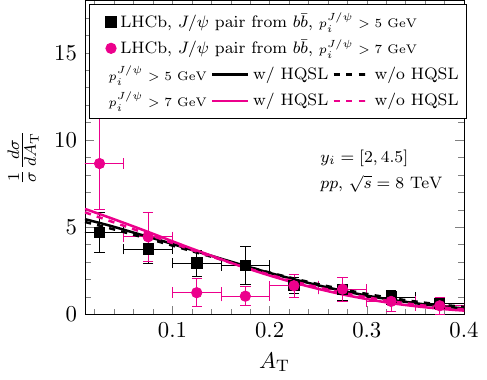}
\caption{
Comparison of our theoretical predictions with LHCb data for the $\Delta\phi$ (upper panel), $\boldsymbol{q}_h$ (middle panel), and $A_{\rm T}$ (lower panel) distributions of $J/\psi$ meson pairs from $b$-hadron decays in forward $pp$ collisions~\cite{LHCb:2017bvf}, illustrating the impact of the single-logarithmic Sudakov contribution from heavy quarks.
}
\label{fig:Jpsi_HQSL}
\end{figure}

\begin{figure}[!ht]
\includegraphics[width=0.7\linewidth]{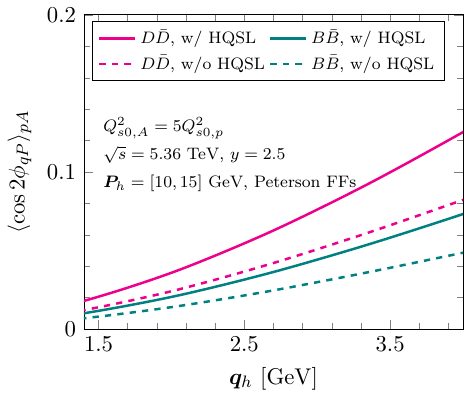}
\caption{Predictions for the $\langle\cos 2\phi_{qP}\rangle_{pA}$ asymmetry as a function of $q_h$ for $D\bar{D}$ pairs (a) and $B\bar{B}$ pairs (b) in $p\text{Pb}$ collisions at $\sqrt{s} = 5.36~\text{TeV}$. Results are shown with and without the single-logarithmic Sudakov contribution from heavy quarks. The calculations are performed at $y = 2.5$ with $P_h \in [10, 15]~\text{GeV}$.}
\label{fig:5360_cos2phi_HQSL}
\end{figure}

\begin{figure}[!ht]\includegraphics[width=0.7\linewidth]{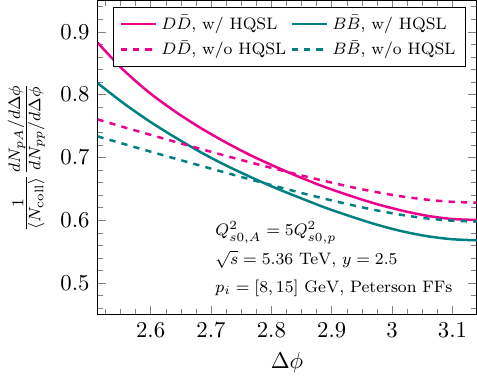}
\includegraphics[width=0.7\linewidth]{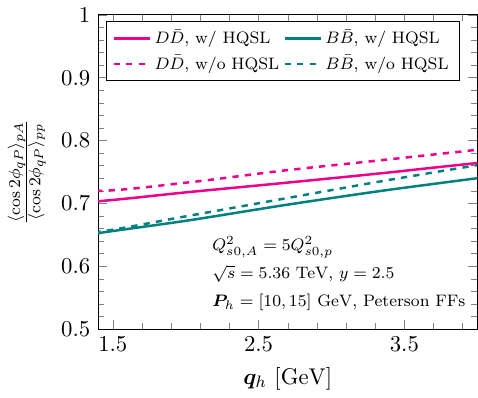}
\caption{Predictions for the $R_{pA}$ (upper panel) and $\langle\cos 2\phi_{qP}\rangle_{pA}/\langle\cos 2\phi_{qP}\rangle_{pp}$ ratio (lower panel)  for $D\bar{D}$ pairs and $B\bar{B}$ pairs at $\sqrt{s} = 5.36~\text{TeV}$ with $y_i = 2.5$. Results are shown with and without the single-logarithmic Sudakov contribution from heavy quarks. }
\label{fig:5360_ratio_HQSL}
\end{figure}

In this subsection, we study the impact of the final-state heavy-quark Sudakov single-logarithmic contribution based on Eq.~(\ref{eq::DDbar}), using the Peterson FFs. The final-state heavy-quark Sudakov single-logarithmic contribution corresponds to the last term of Eq.~(\ref{eq::QSu}). Unlike the KKKS/KKSS FFs, the Peterson FFs are independent of $\mu_F$ and do not resum the collinear single logarithms. Therefore, the impact of the Sudakov single-logarithmic contribution can be assessed more directly.

As shown in Fig.~\ref{fig:dphi_DD_HQSL}, we compare two sets of predictions, with and without the final-state heavy-quark Sudakov single-logarithmic contribution, with the LHCb data for the $D^0\bar{D}^0$ azimuthal-separation ($\Delta\phi$) distributions in $pp$~\cite{LHCb:2012aiv} (upper panel) and $p\text{Pb}$~\cite{LHCb:2020jse} (lower panel) collisions. In addition, in Fig.~\ref{fig:Jpsi_HQSL}, we perform a similar comparison with the LHCb data for $J/\psi$-pair production from $b$-hadron decays in forward $pp$ collisions~\cite{LHCb:2017bvf}, considering the $\Delta\phi$, $\boldsymbol{q}_h$, and $A_{\rm T}$ distributions.

We find that the distributions become noticeably steeper after including the final-state heavy-quark Sudakov single-logarithmic contribution. Nevertheless, the predictions remain in good agreement with the LHCb data for both $D^0\bar{D}^0$ and $J/\psi$-pair production across the different observables considered in $pp$ and $p\text{Pb}$ collisions.

In addition, as shown in Fig.~\ref{fig:5360_cos2phi_HQSL}, we investigate the impact of the heavy-quark single-logarithmic contribution on the $\langle\cos 2\phi_{qP}\rangle$ asymmetry in forward $p\text{Pb}$ collisions at $\sqrt{s}=5.36$ TeV, with $\boldsymbol{P}_h\in[10,15]$ GeV and rapidity $y_i=2.5$. For both $D\bar D$ and $B\bar B$ pair correlations, we find that the predicted asymmetry is larger when the heavy-quark single-logarithmic contribution is included.

Moreover, in Fig.~\ref{fig:5360_ratio_HQSL}, we investigate its impact on $R_{pA}$ (upper panel) and the ratio $\langle\cos 2\phi_{qP}\rangle_{pA}/\langle\cos 2\phi_{qP}\rangle_{pp}$ (lower panel) at $\sqrt{s}=5.36$ TeV and rapidity $y_i=2.5$. For both $D\bar D$ and $B\bar B$ pair correlations, the saturation-induced suppression becomes more pronounced when the heavy-quark single-logarithmic contribution is included. In addition, both observables exhibit the same mass ordering both with and without the heavy-quark single-logarithmic contribution, with stronger suppression for bottom quarks than for charm quarks: $R_{pA}|_{m_b}<R_{pA}|_{m_c}$ and $\frac{\langle \cos 2\phi_{qP} \rangle_{pA}}{\langle \cos 2\phi_{qP} \rangle_{pp}}\big|_{m_b}<\frac{\langle \cos 2\phi_{qP} \rangle_{pA}}{\langle \cos 2\phi_{qP} \rangle_{pp}}\big|_{m_c}$.

\subsection{Dependence on the Saturation Scale $Q_{sA}^2$}
\label{subsec::Qs}

\begin{figure}[!ht]
\includegraphics[width=0.7\linewidth]{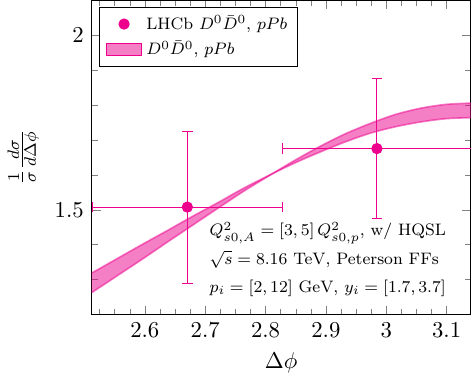}
\includegraphics[width=0.7\linewidth]{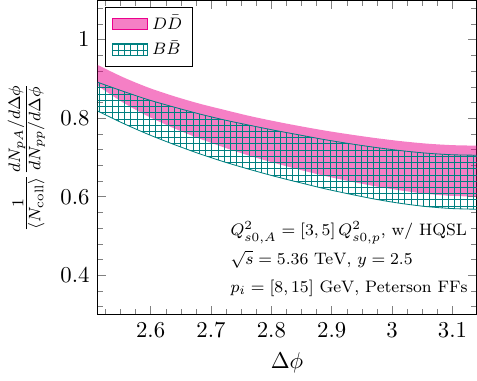}
\caption{
The normalized $\Delta\phi$ distribution and the yield ratio for $D^0\bar{D}^0$ pairs in $p\text{Pb}$ collisions, calculated using the Peterson FFs. The bands correspond to $Q_{s0,A}^2 = [3, 5]Q_{s0,p}^2$.
}
\label{fig:dphi_Qs0}
\end{figure} 

\begin{figure}[!ht]
\includegraphics[width=0.7\linewidth]{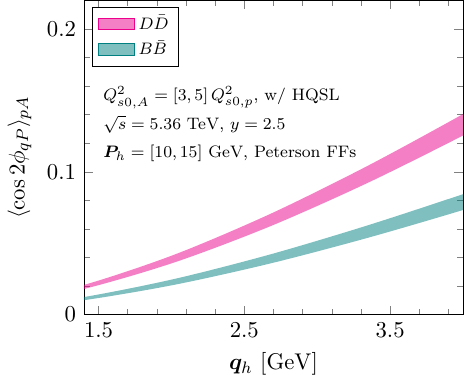}
\includegraphics[width=0.7\linewidth]{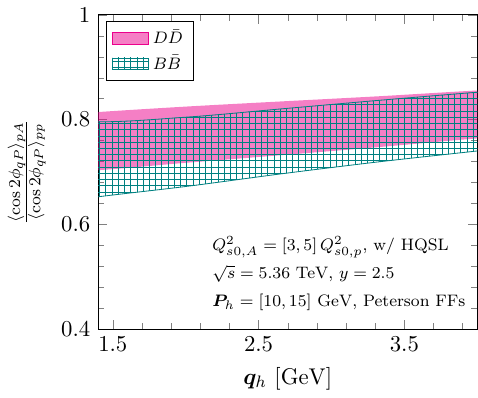}
\caption{The $\langle\cos 2\phi_{qP}\rangle_{pA}$ asymmetries and their ratios between $pA$ and $pp$ collisions as a function of $\bm{q}_h$ for $D\bar{D}$ and $B\bar{B}$ pairs, obtained with the Peterson FFs. The bands indicate the uncertainties from varying $Q_{s0,A}^2 = [3, 5] Q_{s0,p}^2$.
}
\label{fig:cos2phi_Qs0}
\end{figure} 
 
In this section, we discuss the dependence of both the $\Delta \phi$ distribution and $\langle\cos 2\phi_{qP}\rangle_{pA}$ asymmetry on the saturation scale $Q_{sA}^2$. For simplicity, we use only the Peterson FFs in this calculation as an example to show the dependence of predictions on the saturation scale by varying $Q_{s0,A}^2= [3\,Q_{s0,p}^2, 5\,Q_{s0,p}^2]$. 

Fig.~\ref{fig:dphi_Qs0} shows the effect of varying $Q_{s0,A}^2$ on the normalized $\Delta\phi$ distribution and the yield ratio. The bands representing the uncertainties for saturation scale within $Q_{s0,A}^2 = [3, 5] Q_{s0,p}^2$. As shown in the upper panel, the prediction obtained by varying the saturation scale within $Q_{s0,A}^2 = [3\,Q_{s0,p}^2, 5\,Q_{s0,p}^2]$ shows a small uncertainty band and can describe the LHCb data in forward $pA$ collisions. This is because the dominant Sudakov effect hides this saturation broadening effect, resulting in a tiny change in the normalized $\Delta\phi$ distribution. On the other hand, a weaker saturation effect reduces the suppression in $pA$ collisions, leading to a larger $R_{pA}$, as shown in the lower panel of Fig.~\ref{fig:dphi_Qs0}.

Similarly, in Fig.~\ref{fig:cos2phi_Qs0}, we present the influence of $Q_{s0,A}$ on $\langle\cos 2\phi_{qP}\rangle$ and its ratio. As shown in the upper panel, $\langle\cos 2\phi_{qP}\rangle_{pA}$ asymmetry increases as the value of saturation scale $Q_s^2$ decreases. Since the dominant contribution originates from the Sudakov effect rather than the saturation effects, $\langle\cos 2\phi_{qP}\rangle_{pA}$ shows a weak dependence on $Q_s^2$.  In addition, as shown in the lower panel, it gradually approaches unity when $Q_s^2$ decreases, reflecting the weakening of the saturation effect.

In the deep back-to-back region, both $R_{pA}$ and the ratio $\langle\cos(2\phi_{qP})\rangle_{pA}/\langle\cos(2\phi_{qP})\rangle_{pp}$ exhibit the same mass hierarchy, with smaller values for bottom than for charm. The magnitude of this mass dependence is sensitive to the initial saturation-scale ratio $Q_{s0,A}^2/Q_{s0,p}^2$. As this ratio increases, the separation between the charm and bottom results becomes more pronounced, while the mass ordering remains unchanged. For example, as shown in Figs.~\ref{fig:dphi_Qs0} and~\ref{fig:cos2phi_Qs0}, increasing $Q_{s0,A}^2/Q_{s0,p}^2$ from $3$ to $5$ enhances the difference between the charm and bottom results for both the asymmetry ratio and $R_{pA}$.

\section{Conclusion} 
\label{sec:con}

In this work, we have investigated the heavy meson pair production in $pA$ collisions in the back-to-back correlation limit ($\boldsymbol{P}_h\gg\boldsymbol{q}_h$) by employing Sudakov resummation within the CGC formalism.  Our calculations are in good agreement with the $\Delta \phi$ distributions measured by the LHCb Collaboration for $D^0\bar D^0$ pairs in $pp$ and $pA$ collisions. They also agree with the LHCb data on the $\Delta \phi$, $A_T$, and $\boldsymbol{q}_h$ distributions for $J/\psi$ pairs from $b\bar b$ decays in forward $pp$ collisions.  These results demonstrate that our unified resummation framework developed in Ref.~~\cite{Marquet:2025jdr} can consistently describe different experimental measurements.

By comparing the results of $pA$ collisions and $pp$ collisions, the $\Delta \phi$ distribution shows a visible $p_\perp$ broadening effect in $pA$ collisions, while the nuclear modification factor $R_{pA}$ shows a strong saturation-induced suppression at small-$x$. We further present predictions of both $D\bar D$ and $B \bar B$ in the forward rapidity regions at the LHC.  We find that $R_{pA}$ shows a pronounced mass-ordering effect, $R_{pA}\big|_{m_b}<R_{pA} \big|_{m_c}$, consistent with our previous findings in heavy-meson pair photo-production~\cite{Marquet:2025jdr}. In addition, we present predictions for the azimuthal asymmetry $\langle \cos 2\phi_{qP} \rangle$. We find that it receives sizable contributions for both $D$ and $B$ meson pairs together with a clear mass-ordering effect.  The $\langle \cos 2\phi_{qP} \rangle$ ratios between $pA$ and $pp$ collisions also show strong sensitivity to saturation effects.  For both these $\Delta \phi$ and $\langle \cos 2\phi_{qP} \rangle$ measurements, as the rapidity increases,  the suppression both becomes more pronounced. Our work shows that heavy-meson pair correlations in forward $pA$ collisions serve as a clean channel for probing the saturation effect.

\begin{acknowledgements} 
This work was supported in part by the Ministry of Science and Technology of China under Grant No. 2024YFA1611004, by the Natural Science Foundation of Guangdong Province under Grant No.~2026A1515011242, by the CUHK (Shenzhen) University Development Fund under Grant No.~UDF01001859, and by the National Natural Science Foundations of China under Grant No.~12605155.
\end{acknowledgements}

\bibliographystyle{apsrev4-1}
\bibliography{main.bib}

\end{document}